\documentclass[prd,superscriptaddress,amsfonts,amssymb,amsmath,showpacs]{revtex4-2}
\usepackage{bm}
\usepackage{amsfonts}
\usepackage{latexsym}
\usepackage{graphicx}
\usepackage{amsmath}
\usepackage{palatino}
\usepackage{mathpazo}
\usepackage{natbib}
\usepackage{textcomp}
\usepackage{booktabs}
\usepackage{dcolumn}
\usepackage{booktabs}
\usepackage{multirow}
\usepackage{hyperref}
\hypersetup{colorlinks,citecolor=blue}
\usepackage{amsmath}
\usepackage{xcolor}
\usepackage{orcidlink}
\usepackage[caption=false]{subfig}
\usepackage{commath}
\usepackage{lmodern}

\usepackage{anyfontsize}
\usepackage{amssymb}
\usepackage{amsthm}
\usepackage{amsmath}
\usepackage{lipsum}
\usepackage{float}
\usepackage{orcidlink}

\def\jnl@style{\it}
\def\aaref@jnl#1{{\jnl@style#1}}

\def\aaref@jnl#1{{\jnl@style#1}}

\def\aj{\aaref@jnl{AJ}}                   
\def\apj{\aaref@jnl{ApJ}}                 
\def\apjl{\aaref@jnl{ApJ}}                
\def\apjs{\aaref@jnl{ApJS}}               
\def\apss{\aaref@jnl{Ap\&SS}}             
\def\aap{\aaref@jnl{A\&A}}                
\def\aapr{\aaref@jnl{A\&A~Rev.}}          
\def\aaps{\aaref@jnl{A\&AS}}              
\def\mnras{\aaref@jnl{Mon.~Not.~Roy.~Astron.~Soc.}}             
\def\prd{\aaref@jnl{Phys.~Rev.~D}}        
\def\prc{\aaref@jnl{Phys.~Rev.~C}}  
\def\prl{\aaref@jnl{Phys.~Rev.~Lett.}}    
\def\qjras{\aaref@jnl{QJRAS}}             
\def\skytel{\aaref@jnl{S\&T}}             
\def\ssr{\aaref@jnl{Space~Sci.~Rev.}}     
\def\zap{\aaref@jnl{ZAp}}                 
\def\nat{\aaref@jnl{Nature}}              
\def\aplett{\aaref@jnl{Astrophys.~Lett.}} 
\def\apspr{\aaref@jnl{Astrophys.~Space~Phys.~Res.}} 
\def\physrep{\aaref@jnl{Phys.~Rep.}}      
\def\physscr{\aaref@jnl{Phys.~Scr}}       
\def\commat{\aaref@jnl{Comm.~Math.~Phys.}}              
\def\science{\aaref@jnl{Science}}               
\def\cqg{\aaref@jnl{Classical Quant.~Grav.}}            
\def\jpcs{\aaref@jnl{JPCS}}                                     
\def\ijmpd{\aaref@jnl{Int.~J.~Mod.~Phys.~D}}                    
\def\grg{\aaref@jnl{Gen.~Relat.~Gravit.}}               
\def\rpp{\aaref@jnl{Rep.~Prog.~Phys.}}          
\def\npa{\aaref@jnl{Nucl.~Phys.~A}}        
\def\lrr{\aaref@jnl{Living Rev.~Rel.}}                   
\def\jcap{\aaref@jnl{J.~Cosmology Astropart.~Phys.}}    
\def\rmp{\aaref@jnl{Rev.~Mod.~Phys.}}   
\def\epjc{\aaref@jnl{Eur.~Phys.~J.~C}}

\allowdisplaybreaks[1]
\renewcommand{\arraystretch}{1.1}
\begin{document}

\color{black}       

\title{Bouncing Cosmology and Cosmological Dynamics in $f(Q,T)$ Gravity}

\author{Bhagwat Gidhad}
\email{bbg.scos@jspmuni.ac.in}
\affiliation{Department of Mathematics, School of Computational Sciences, JSPM University Pune-412207, India.}
\author{A. S. Agrawal\orcidlink{0000-0003-4976-8769}}
\email{asagrawal.sbas@jspmuni.ac.in\\
agrawalamar61@gmail.com}
\affiliation{Department of Mathematics, School of Computational Sciences, JSPM University Pune-412207, India.}
\author{S. A. Kadam\orcidlink{0000-0002-2799-7870}}
\email{siddheshwar.kadam@dypiu.ac.in;
\\k.siddheshwar47@gmail.com}
\affiliation{Centre for Interdisciplinary Studies and Research, D Y Patil International University, Akurdi, Pune-411044, Maharashtra, India.}
\date{\today}

\begin{abstract}
\textbf{Abstract}: 
We propose a reconstructed cosmological model in the framework of $f(Q,T)$ gravity, that provides a unified description of the early- and late-time evolution of the Universe. The model exhibits a non-singular asymmetric bounce, smoothly connecting an initial contracting phase to the subsequent expanding Universe and naturally evolving into a late-time dark energy-dominated epoch. Our study focuses on the progression of the Hubble parameter, energy density, pressure, and the  parameter. This analysis aims to define the various stages of cosmic evolution and explore the characteristics of dark energy. The analysis of energy conditions reveals that the essential conditions for achieving a non-singular bounce are violated. Overall, the $f(Q, T)$ gravity model, once reconstructed, effectively captures the cosmic dynamics surrounding the bounce. It offers a cohesive theoretical framework that reliably explains the Universe's evolution during both its early and late stages.

\end{abstract}

\maketitle
\section{Introduction}
Modern cosmology is fundamentally based on Einstein's General Relativity (GR), which has successfully explained the gravitational interaction through the curvature of spacetime described by Riemannian geometry\cite{einstein1922general}. Within this framework, the inflationary paradigm has emerged as the standard description of the very early Universe, providing elegant solutions to the flatness, monopole, and monopole challenges while at the same time explaining the origin of primordial density perturbations and gravitational waves \cite{Guth-1981-23, Carroll-2001-4, Weinberg2008cosmology, LINDE1982389, Peter-2002, Farrugia-2018-97}. Motivated by the remarkable agreement between theoretical predictions and cosmological observations, including Type Ia supernovae and the cosmic microwave background, considerable efforts have been devoted to understanding the complete evolutionary history of the Universe within a unified theoretical framework \cite{Riess-1998-116, Perlmutter-1998-517, di2025cosmoverse}.

Despite its remarkable success, the inflationary scenario does not resolve the problem of the initial cosmological singularity. This limitation has motivated the development of alternative early-Universe models, among which bouncing cosmology has attracted significant attention. In a bouncing Universe, the present expansion is preceded by a contracting phase, thereby taking the place of the Big Bang singularity with a smooth non-singular bounce \cite{Ashtekar-2006-74, Odintosv-2016}. Consequently, bouncing cosmology provides a natural framework for avoiding the initial singularity while maintaining a consistent description of cosmic evolution \cite{lohakare2022bouncing, agrawal2022bouncing, gul2024comprehensive, singh2023bouncing, Bamba-2014-01}. At the bounce point $(t_b)$, the Hubble parameter satisfies $H(t_b)=0$, and changes continuously from negative to positive values
($H<0 ~ \rightarrow ~ H>0$),
indicating the change from a contracting phase to an expanding Universe \cite{Cai-2007-2007, Shabani-2018}. Bouncing cosmologies have therefore been extensively investigated in GR and several modified theories of gravity, including teleparallel and symmetric teleparallel gravity \cite{bamba2014bouncing, Singh-2023, kadam2026bouncing, mishra2024scalar, Caruana-2020, delaCruz-Dombriz-2018-97}.

The search for viable gravitational theories beyond GR has also motivated the exploration of non-Riemannian geometries. The earliest attempt was proposed by Weyl, who generalized Riemannian geometry by introducing non-metricity in an effort to unify gravitation and electromagnetism \cite{Weyl-1918}. Although Einstein pointed out inconsistencies of Weyl's original theory with physical observations \cite{Einstein-1918}, the concept of non-metricity remained an important geometrical ingredient. Subsequently, Cartan extended GR by incorporating torsion, leading to the Einstein--Cartan theory \cite{Cartan-1923}, while Weitzenböck introduced a curvature-free spacetime characterized by non-vanishing torsion, laying the foundation of teleparallel gravity \cite{Weitzenbock-1923}. Einstein later utilized this geometry to formulate a unified teleparallel description of gravitation and electromagnetism, which eventually evolved into the Teleparallel Equivalent of General Relativity (TEGR), where gravity is described entirely by torsion instead of curvature \cite{Einstein-1928-17, Pellegrini-1963}.

These developments reveal that GR admits three equivalent geometrical descriptions: (i) the curvature formulation based on the Ricci scalar, (ii) the teleparallel formulation depending on the torsion scalar $\mathcal{T}$, and (iii) the symmetric teleparallel formulation based on the non-metricity scalar $Q$, where both curvature and torsion vanish \cite{Nester-1999,gadbail2021power}. The formulation at a later stage has recently attracted considerable attention due to its simple geometrical structure and promising cosmological implications. A natural extension of symmetric teleparallel gravity is $f(Q)$ gravity \cite{jimenez2018coincident}, which has subsequently been generalised to $f(Q, T)$ gravity by introducing a non-minimal coupling between the non-metricity scalar $Q$ and the trace of the energy--momentum tensor $T$ \cite{Xu19}. $f(Q)$ gravity does not suffer from strong coupling issues in cosmological perturbations around the Friedmann--Lema\^{\i}tre--Robertson--Walker (FLRW) background, making it an attractive framework for studying the evolution of the Universe \cite{Lu-2019, Dialektopoulos-2019, Bajardi-2020, Jimenez-2020, Wu-2010, Li-2011, Cai-2016, Benetti-2021, Golovnev-2018}. Furthermore, extensive studies have demonstrated the cosmological viability of these theories, including linear perturbations, gravitational waves, and late-time cosmic acceleration \cite{Frusciante-2021, Latorre-2018, Soudi-2019, Zia-2021, Pati-2021}.

Motivated by the success of modified gravity in addressing both the early- and late-time cosmic development, bouncing cosmological models have been widely investigated in several extended gravitational theories, including $f(R)$, $f(R,T)$, $f(R,G)$, $f(Q,T)$, $f(\mathcal{T})$, and $f(\mathcal{T},B)$ gravity \cite{Bhattacharya-2016, Saridakis-2018, ilyas2021bounce, Shabani-2018, Mishra-2019-34, Tripathy-2021-71, Elizalde-2020-954, Logbo-2019, skugoreva2020bouncing, AGRAWAL2021100863,ilyas2024gravastars}. Comprehensive reviews of bouncing cosmology and its observational implications, including possible signatures in the cosmic microwave background, may be found in Refs.~\cite{Novello:2008ra, Battefeld:2014uga, Liu-2013}. These investigations demonstrate that both modifications of the gravitational action and exotic matter sources can successfully realize a non-singular bouncing Universe \cite{Boisseau-2015}. Motivated by these developments, the present work investigates bouncing cosmology within the basic structure of $f(Q,T)$ gravity, where the gravitational interaction is governed by the non-metricity scalar $Q$ together with its non-minimal coupling to the trace of the energy--momentum tensor $T$ \cite{myrzakulov2023constraining}. The objective is to examine whether this geometric framework can naturally accommodate a viable non-singular bouncing scenario while remaining consistent with the fundamental requirements of cosmological evolution.

The organization of this paper is as follows. In Sec.~\ref{sec.2}, we outline the fundamental formalism of field equations in $f(Q,T)$ gravity. In Sec.~\ref{sec.3}, we analyze the evolution of the scale factor, the Hubble parameter, the effective equation of state (EoS) parameter, and the comoving Hubble radius. The evolution of the cosmological quantities, namely the pressure, energy density, and the  parameter, is discussed in Sec.~\ref{sec4}. To examine the physical viability of the model, we investigate the energy conditions, namely the null energy condition (NEC), strong energy condition (SEC), and dominant energy condition (DEC), in Sec.~\ref{sec.5}. Finally, the main conclusions of this work are presented in Sec.~\ref{sec.6}.

\section{Basic formalism of field equation of $f(Q,T)$ gravity theory}\label{sec.2}

The action equation of $f(Q,T)$ gravity \cite{Xu19,bhagat2026effects} is,
\begin{equation} \label{eq.1}
S=\int\left(\dfrac{1}{16\pi}f(Q,T)+\mathcal{L}_{m}\right)d^{4}x\sqrt{-g},
\end{equation}

where $\mathcal{L}_{m}$ denotes the matter Lagrangian density, and $g=\det[g_{\mu\nu}]$ is the determinant of the metric tensor $g_{\mu\nu}$. Here, $Q$ represents the non-metricity scalar, while $T=g^{\mu \nu}T_{\mu \nu}$ denotes the trace of the energy--momentum tensor.
Varying the gravitational action \eqref{eq.1}, the field equation of $f(Q,T)$ gravity can be obtained as, 

\begin{equation}\label{eq.2}
-\frac{2}{\sqrt{-g}}\bigtriangledown_{k}(f_{Q}\sqrt{-g}P^{k}_ {\mu \nu})-\frac{1}{2}fg_{\mu \nu}+f_{T}(T_{\mu \nu}+\Theta_{\mu \nu})-f_{Q}(P_{\mu kl} Q^{\;\;\; kl}_{\nu}-2Q^{kl}_{\;\;\;\mu} P_{kl\nu})=8 \pi T_{\mu \nu},
\end{equation}

where $f_Q=\frac{\partial f(Q,T)}{\partial Q}$, $\Theta_{\mu \nu}\equiv g^{\alpha \beta}\frac{\delta T_{\alpha \beta}}{\delta g^{\mu \nu}}$ and the non-metricity scalar $Q$ is described as,
\begin{equation} \label{eq.3}
Q\equiv -g^{\mu \nu}( L^k_{~l\mu}L^l_{~\nu k}-L^k_{~lk}L^l_{~\mu \nu}) , 
\end{equation}

where,  $L^k_{~l\gamma}\equiv -\frac{1}{2}g^{k\lambda}(\bigtriangledown_{\gamma}g_{l\lambda}+\bigtriangledown_{l}g_{\lambda \gamma}-\bigtriangledown_{\lambda}g_{l\gamma})$. 

The energy-momentum tensor is, 
\begin{equation}
    T_{\mu \nu}\equiv-\frac{2}{\sqrt{-g}}\frac{\delta(\sqrt{-g}\mathcal{L}_m)}{\delta g^{\mu \nu}},
\end{equation}

The model super potential can be represented as,
\begin{equation} \label{eq.4}
 P^{k}_{\mu \nu}=-\frac{1}{2}L^{k}_{\mu \nu}+\frac{1}{4}(Q^{k}-\tilde{Q}^{k})g_{\mu \nu}-\frac{1}{4}\delta^{k}_{(\mu}Q_{\nu)}.   
\end{equation}

The trace of the energy-momentum tensor and the trace of the non-metricity tensor can be respectively denoted as,
\begin{eqnarray}
T= T_{\mu \nu}g^{\mu \nu}, \quad 
Q_{k}= Q_{k}^{\;\;\mu}\;_{\mu}, \quad \tilde{Q}_{k}=Q^{\mu}\;_{k\mu}. \nonumber
\end{eqnarray}

We adopt the assumption of a homogeneous, isotropic and spatially flat Universe, the FLRW space-time as,
\begin{eqnarray}\label{eq.5}
ds^{2}=-N^{2}(t)dt^{2}+a^{2}(t)(dx^{2}+dy^{2}+dz^{2}),
\end{eqnarray}
where $N(t)$ denote the lapse function and $a(t)$ denote the scale factor. The scale factor provides information about the contracting or expanding behaviour of the Universe, while the rate of expansion or contraction is characterized by the Hubble parameter. 

The dilation rate and non-metricity scalar are given by 
\begin{equation} 
\tilde{T}=\frac{\dot{N}(t)}{N(t)}, \quad  Q = 6\frac{H^2}{N^2}. 
\end{equation} 
where an overdot represents the derivative with respect to time $t$. 
The Universe's evolution depends on the scale factor $a(t)$ which tells whether the Universe is expanding or contracting.
The Hubble parameter is used to measure the rate of expansion, and it is defined as
$H(t)=\dot{a}(t)/a(t)$. For the spatially flat FLRW metric with the lapse function $N(t)=1$. 

The energy-momentum tensor for the perfect fluid is expressed as,
\begin{equation}
T^{\mu}_{\ \nu}=\mathrm{diag}(-\rho,p,p,p),    
\end{equation}

where $\rho$ and $p$ are the energy density and pressure, respectively.

Now, the field equations \eqref{eq.2} can be deduced as,
\begin{subequations}
\begin{eqnarray}
p&=&-\frac{1}{16\pi}\left[f-12f_{Q}H^2-4\Xi \right]\label{p1} \\
\rho&=&\frac{1}{16\pi}\left[f-12f_{Q} H^2-4\Xi\left(\frac{\kappa}{\kappa +1}\right)\right]\label{rho1},
\end{eqnarray}
\end{subequations}

where $\Xi =\frac{d}{dt}[f_{Q}H]$ and $8\pi \kappa\equiv f_{T}=\frac{\partial f}{\partial T}$. 

Adding eqns. \eqref{p1} and \eqref{rho1}, the evolution equation for $\Xi$ can be obtained as,
\begin{equation} \label{eq.8}
\Xi=4\pi (\kappa +1)(p+\rho).
\end{equation}

Considering the general forms for $f(Q,T)$, such as  $f(Q,T)=\alpha Q^{m}+\beta T$ \cite{Xu19}. Here, $\alpha$ and $\beta$ are arbitrary constants. For this functional, we have $\alpha = f_{Q}/ (m Q^{m-1})$, $\beta = 8\pi\kappa$. Also we have $\dot{f_{Q}}=2(m-1)f_{Q}\frac{\dot{H}}{H}$ and $\Xi = f_{Q} \dot{H}(2m-1)$. Now, from Eqns. \eqref{p1} and \eqref{rho1}, we obtain the pressure and energy density as, 

\begin{subequations}
\begin{eqnarray}\label{p2}
p&=& \frac{2\Xi[3\kappa +2]-\alpha Q^{m}(\kappa +1)(1-2m)}{16\pi (\kappa +1)(2\kappa +1)},  \\
\rho &=&   \frac{2\Xi \kappa +\alpha Q^{m}(\kappa +1) (1-2m)}{16\pi (\kappa +1)(2\kappa +1)}. \label{rho2}
\end{eqnarray}
\end{subequations}

The equation of state (EoS) parameter can be obtained as,
\begin{equation}
\omega =\frac{p}{\rho } = -1+\frac{4\Xi [2\kappa +1]}{2\Xi \kappa +\alpha Q^{m} (\kappa +1)  (1-2m)}   \label{eq.13}
\end{equation} 

\section{Evolution of the cosmological dynamical parameters}\label{sec.3}

\subsection{Evolution of the scale factor and Hubble parameter}
Most studies on bouncing cosmology have primarily focused on resolving the problem of the initial Big Bang singularity by replacing the singular origin of the Universe with a non-singular bouncing phase. On the other hand, much attention has been devoted to the late-time accelerated expansion of the Universe using a variety of dark energy models. However, comparatively little attention has been given to cosmological models that simultaneously address both the early-time singularity problem and the late-time acceleration of the Universe. Motivated by this, we consider a hybrid scale factor that combines the features of a matter bounce model with those of an exponential expansion model. The matter bounce component ensures a non-singular cosmological evolution by avoiding the initial Big Bang singularity, whereas the exponential component successfully describes the late-time accelerated expansion associated with dark energy. Therefore, the scale factor proposed here is a hybrid scale factor to include the early- and late-time dynamics of the Universe at the same time.

The hybrid scale factor and its corresponding Hubble parameter \cite{odintsov2021unifying} are given by
\begin{subequations}
\begin{eqnarray}
   a(t)&=&\left(a_{0} t^2+1\right)^n e^{\frac{(t_{s} -t)^{1-\gamma }}{\gamma -1}}, \\ 
   H(t)&=&\frac{2 a_{0} n t}{a_{0} t^2+1}+(t_{s} -t)^{-\gamma }.
\end{eqnarray}
\end{subequations}

\begin{figure}[H]
    \centering
    \includegraphics[width=0.45\linewidth]{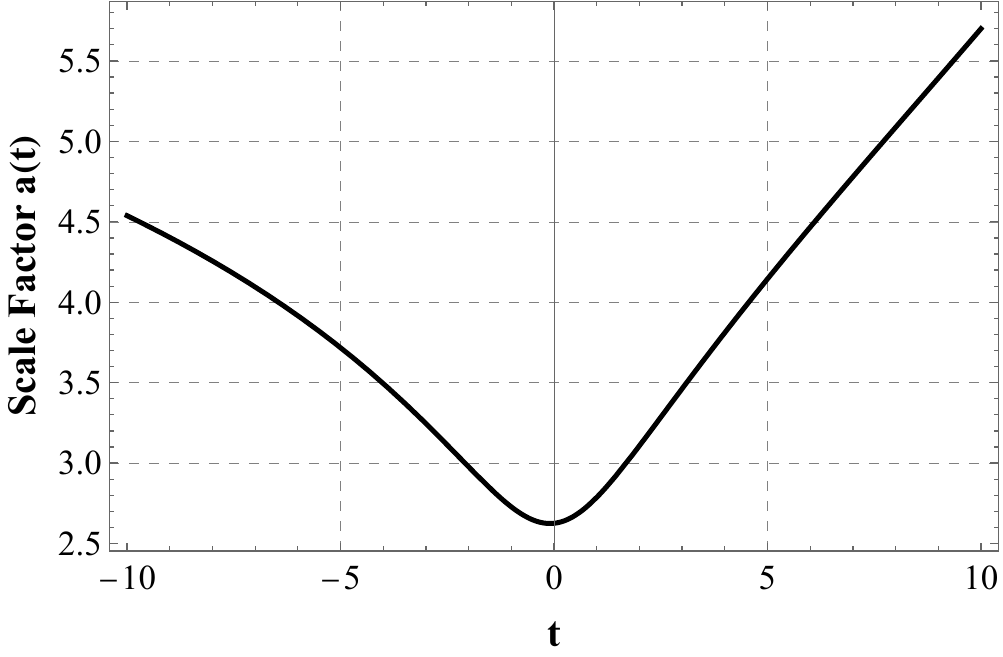}
    \hfill
    \includegraphics[width=0.45\linewidth]{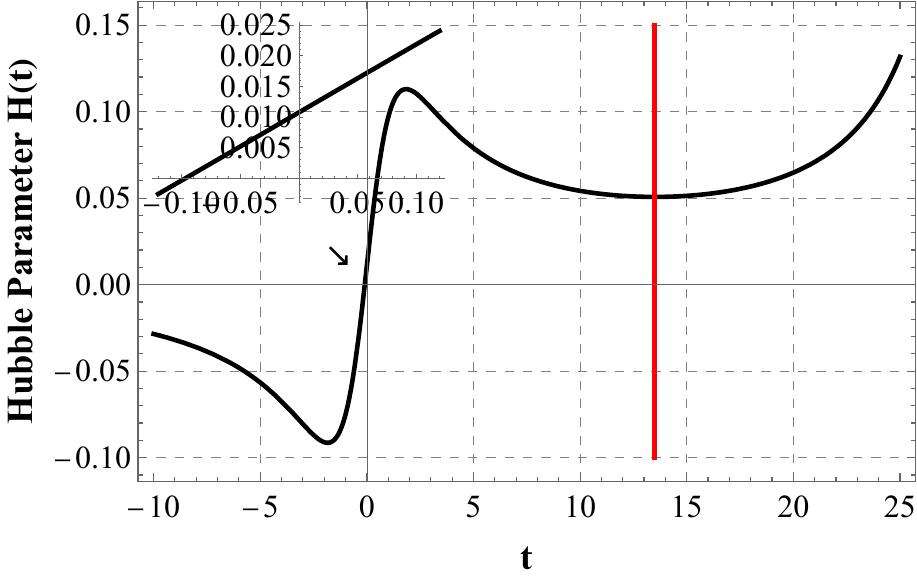}
    
    \caption{Evolution of the scale factor and the Hubble parameter with cosmic time. The left and right panels represent the scale factor and the Hubble parameter, respectively, for the chosen model parameters $a_0=0.3$, $\gamma=\frac{4}{3}$, $n=0.185$, and $t_s=30$.}
    \label{fig:scale_hubble}
\end{figure}

respectively. Their graphical behaviour is illustrated in Fig.~\ref{fig:scale_hubble}. From the figure, it is evident that the Universe undergoes a non-singular bounce at $t=-0.09$, where the scale factor reaches its absolute minimum prior to changing from contracting to expanding. Furthermore, the Hubble parameter changes its sign at the bounce, confirming the occurrence of the matter bounce, while the exponential expansion governs its subsequent evolution at late times.   

\subsection{The Evolution of Effective Equation of State Parameter}
The effective EoS parameter can be represented as
\begin{equation}
    \omega_{\mathrm{eff}}=-1-\frac{2\dot{H}}{3H^{2}}
\end{equation}
\begin{figure}[H]
    \centering   \includegraphics[width=0.45\textwidth]{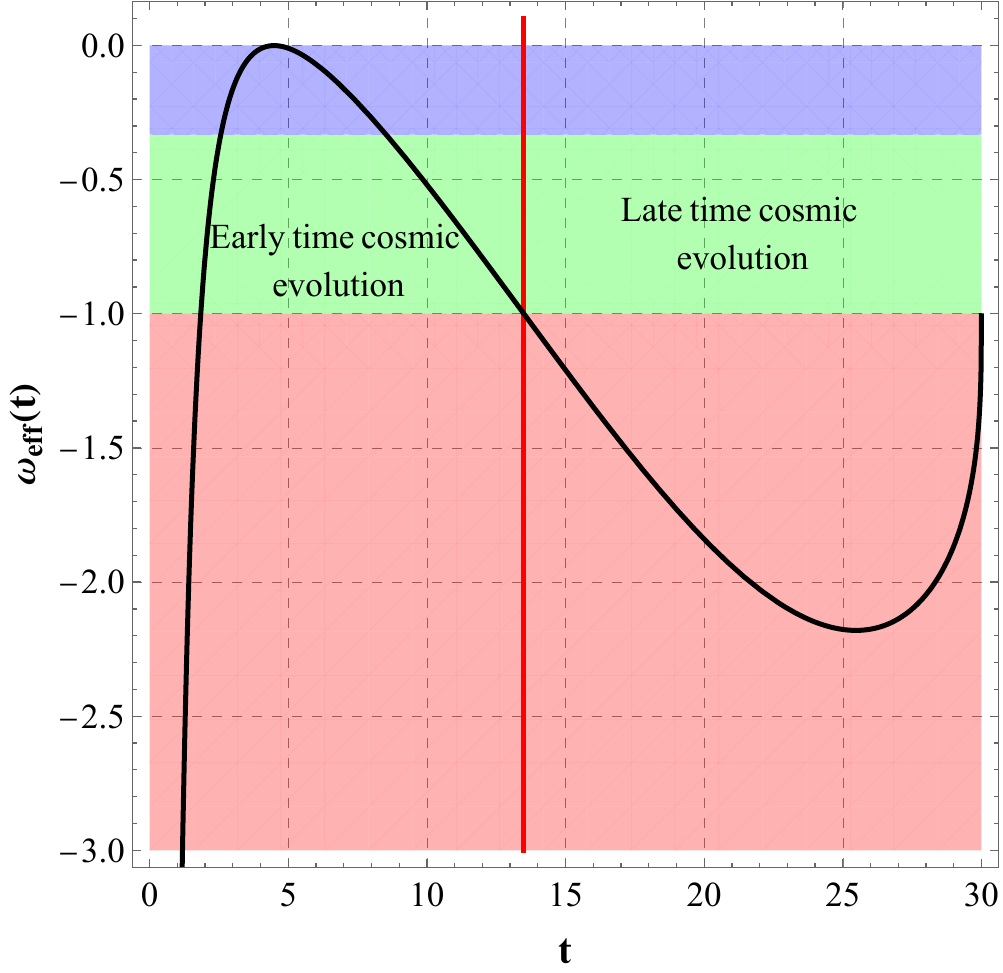}
    \caption{The effective  parameter as functions of cosmic time for the chosen model parameters $a_0=0.3$, $\gamma=\frac{4}{3}$, $n=0.185$, and $t_s=30$.}
    \label{fig:omegaeff}
\end{figure}
The effective equation-of-state (EoS) parameter, $\omega_{\mathrm{eff}}$, provides valuable information about the evolution of the Universe. Different values of $\omega_{\mathrm{eff}}$ correspond to different cosmological epochs and characterize the dynamical behavior of the cosmic fluid. Therefore, analyzing the evolution of $\omega_{\mathrm{eff}}$ offers an effective way to trace the expansion history of the Universe and identify transitions between different evolutionary phases. From Fig.~\ref{fig:omegaeff}, it is noteworthy that the chosen model parameters yield a physically consistent evolution of the effective EoS parameter, $\omega_{\mathrm{eff}}$. The figure clearly illustrates three distinct cosmological regimes. The interval $-1<\omega_{\mathrm{eff}}<-1/3$, shaded in green, corresponds to the accelerated expansion phase of the Universe driven by quintessence-like dark energy. The region $-1/3<\omega_{\mathrm{eff}}<0$, shaded in blue, represents the decelerating phase, where the cosmic dynamics are dominated by ordinary matter or radiation. In contrast, the regime $\omega_{\mathrm{eff}}<-1$, shaded in red, characterizes the phantom phase, which is associated with a super-accelerated expansion of the Universe.

It is observed that, at early cosmic times, the Universe evolves through a matter-dominated epoch and subsequently crosses the $\Lambda$CDM boundary, $\omega_{\mathrm{eff}}=-1$. At the present cosmic age, $t \approx 13.8\,\mathrm{Gyr}$, the effective EoS parameter approaches $\omega_{\mathrm{eff}} \simeq -1$,
indicating an evolution very close to the cosmological constant scenario. This behavior suggests that the present Universe is dominated by dark energy, in good agreement with current cosmological observations. Therefore, the obtained evolution of $\omega_{\mathrm{eff}}$ successfully reproduces the observed transition from a decelerated matter-dominated phase to the present accelerated expansion of the Universe.

\subsection{Comoving Hubble Radius}

The bouncing scenario of the Universe consists of three distinct evolutionary phases: an early-time contracting era, an intermediate bounce phase, and a late-time expanding era. The intermediate stage, known as the \textit{bounce epoch}, marks the transition of the Universe from contraction to expansion, thereby avoiding the initial cosmological singularity predicted by the standard Big Bang model. Over the years, several bouncing cosmological models have been proposed in the literature. Among them, the \textit{matter bounce cosmology} has attracted considerable attention because it provides a nonsingular alternative to the standard cosmological paradigm while naturally generating a nearly scale-invariant spectrum of primordial perturbations.

Although these attractive characteristics are present, the conventional matter bounce model fails to sufficiently explain the observed accelerated expansion of the Universe at later times, which is thought to be caused by dark energy. To overcome this
limitation, we adjust the scale factor by adding an exponential component. This modification allows the model to explain both the nonsingular bouncing behavior in the early Universe and the accelerated expansion at later times within a single framework.\cite{bamba2014bouncing}.

\begin{figure}[H]
    \centering   \includegraphics[width=0.45\textwidth]{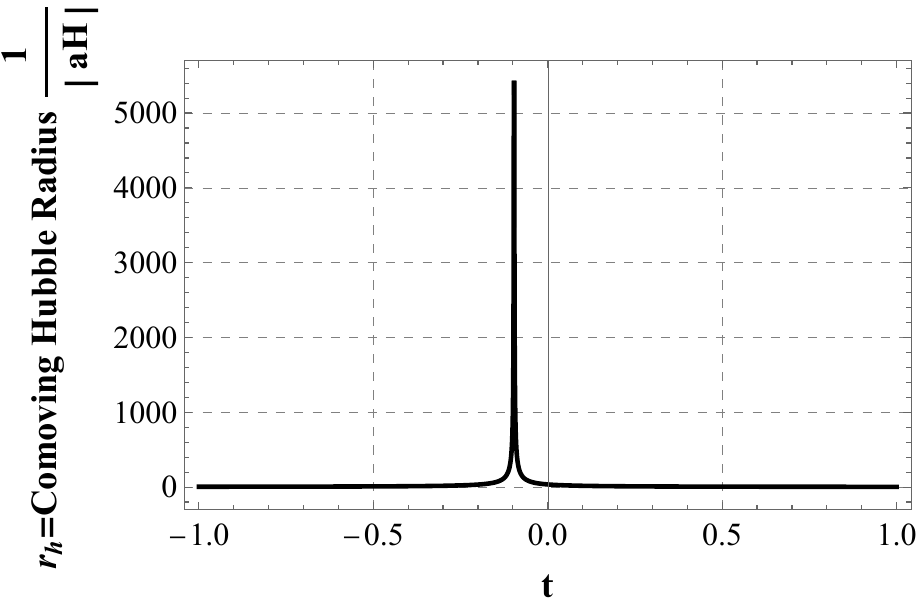}
    \includegraphics[width=0.45\textwidth]{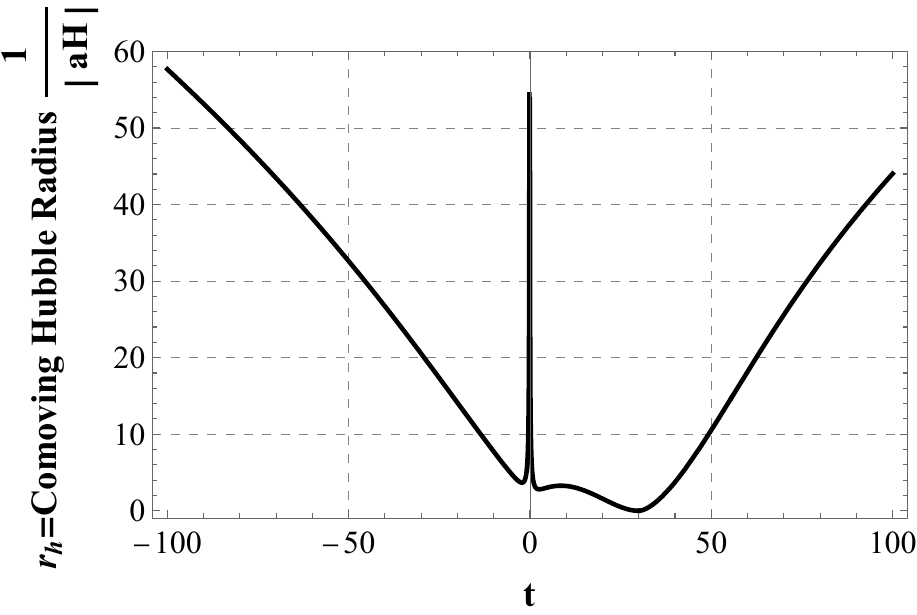}
    \caption{The Comoving Hubble radius as functions of cosmic time for the chosen model parameters $a_0=0.3$, $\gamma=\frac{4}{3}$, $n=0.185$, and $t_s=30$.}
    \label{fig:HR}
\end{figure}

An important feature of a bouncing cosmological model is that the Hubble parameter vanishes, $H=0$, at the bounce epoch, indicating the shift from the Universe's contracting phase to its expanding phase \cite{odintsov2020bottom}. Consequently, the comoving Hubble radius, defined as $r_h=1/(aH)$, diverges because the Hubble parameter at the bounce point becomes zero. This divergence is evident in the left panel of Fig.~\ref{fig:HR}, confirming the characteristic behavior expected in nonsingular bouncing cosmologies. This divergence is a direct consequence of the vanishing Hubble parameter and reflects the reversal of the cosmic evolution from contraction to expansion. The comoving Hubble radius diverges asymptotically (right panel of Fig.~\ref{fig:HR}), corresponding to a late-time decelerating Universe, where the Hubble horizon continuously grows.

\section{Evolution of the Cosmological Parameter}\label{sec4}

\begin{figure}[H]
    \centering
    \includegraphics[width=0.45\textwidth]{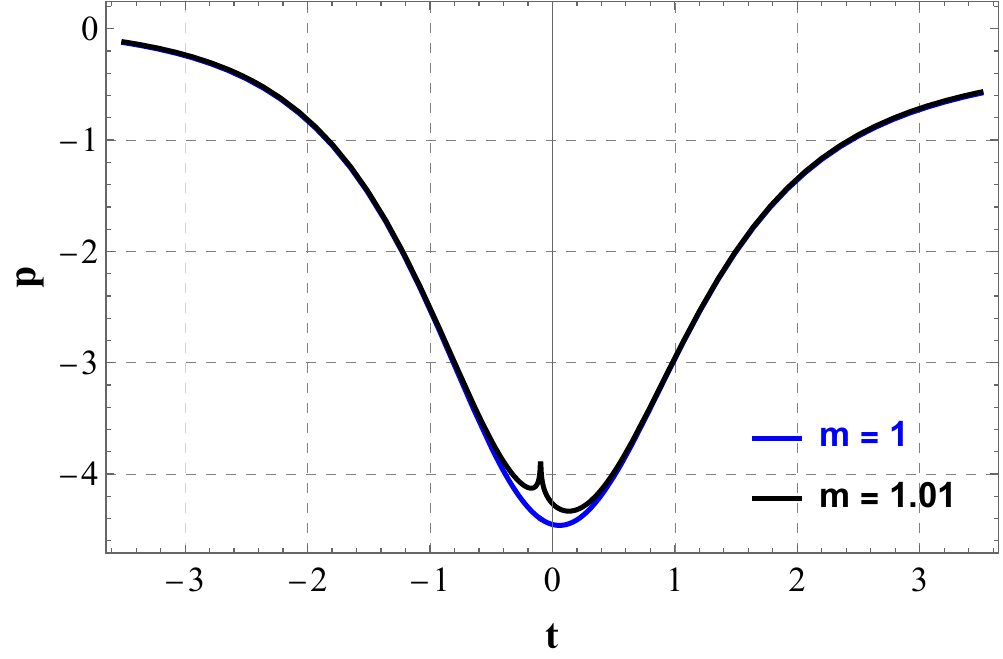}
    \hfill
    \includegraphics[width=0.45\textwidth]{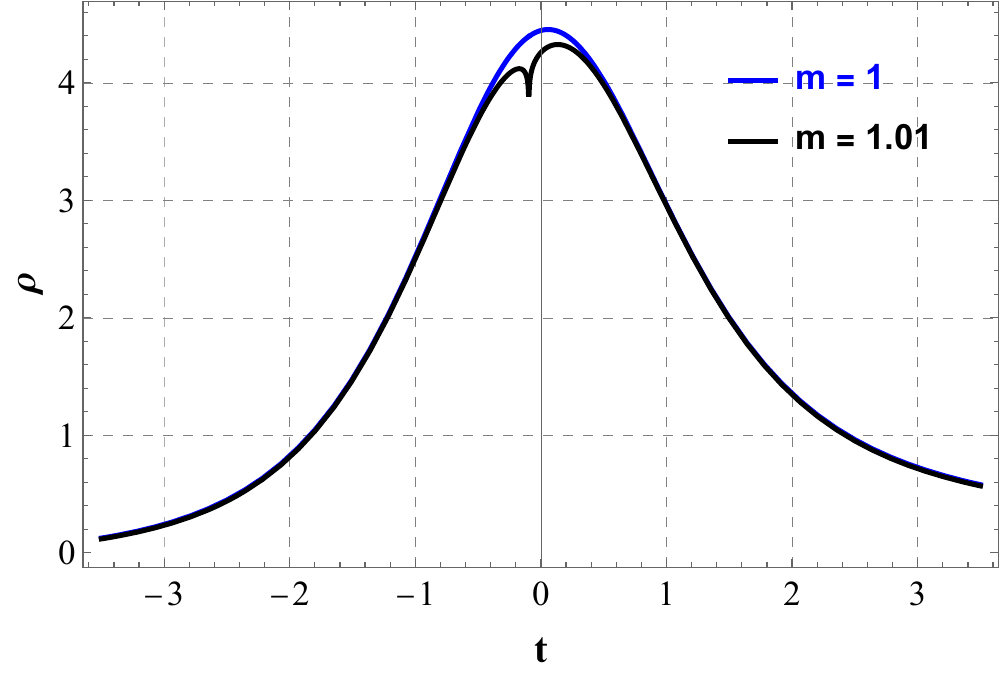}
    \vspace{0.4cm}
    \includegraphics[width=0.45\textwidth]{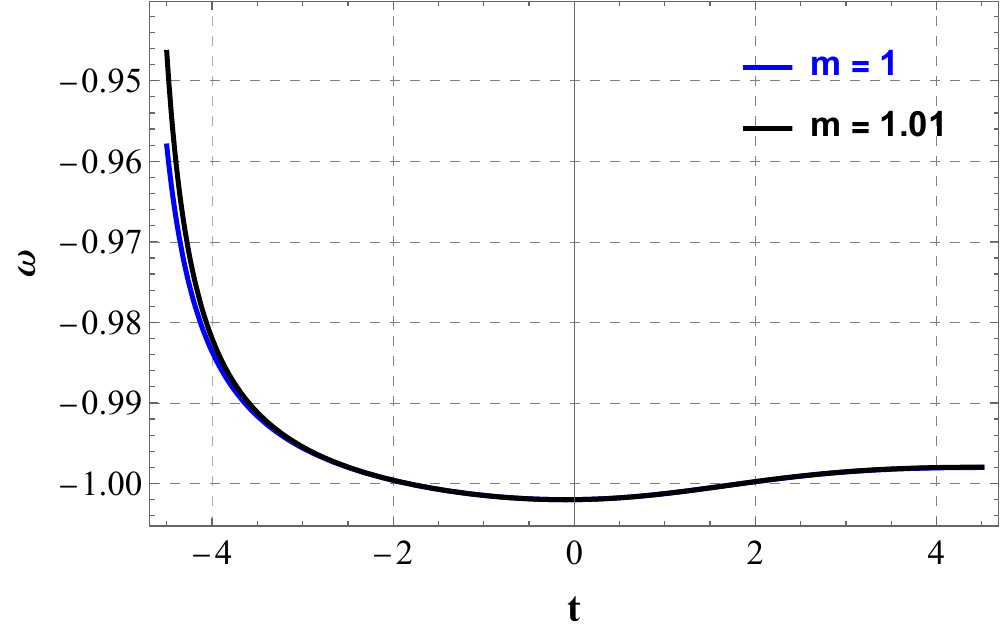}
    \caption{Evolution of the pressure, energy density, and EoS parameter as functions of cosmic time for the chosen model parameters $\kappa=-0.49975$, $\alpha=-0.5$, $a_0=0.3$, $\gamma=\frac{4}{3}$, $n=0.185$, and $t_s=30$.}
    \label{fig:p_rho_omega}
\end{figure}

The expressions for the pressure and energy density in $f(Q,T)$ gravity are given by Eqs.~\eqref{p2} and \eqref{rho2}, respectively. Using the parametric values fixed in the previous section for the analysis of the scale factor and the Hubble parameter, we now investigate the behaviour of the pressure and energy density at and around the bounce epoch. The Fig.~\ref{fig:p_rho_omega} [Left panel] shows how the pressure changed over time. The pressure remains negative throughout the entire cosmic evolution, both before and after the bounce. As the Universe approaches the bounce epoch, the pressure decreases rapidly and attains its minimum value at $t=0$. However, when the power of the non-metricity scalar $Q$ deviates from its linear form in the $f(Q,T)$ model, a small bump appears near the bounce point ($t\approx -0.09$). For the linear dependence on $Q$, this feature is clearly visible; shown in the blue curve. Although the bounce occurs at $t\approx -0.09$, the minimum value of the pressure is attained at $t=0$. The emergence of the bump for nonlinear powers of $Q$ shows that the non-metricity contribution significantly influences the dynamics surrounding the bounce and modifies the evolution of the pressure in its vicinity. 

The right panel of  Fig.~\ref{fig:p_rho_omega} shows the comparable evolution of the energy density. Its behaviour is found to be nearly complementary to that of the pressure \cite{mishra2021dynamics}. The energy density rises as the Universe gets closer to the bounce, reaching its maximum value at $t=0$, and subsequently decreases as the Universe evolves away from the bounce. Similar to the pressure profile, a slight bump is also observed near $t\approx -0.09$ when nonlinear powers of the non-metricity scalar are considered. This behaviour further supports the conclusion that higher-order contributions of $Q$ have an impact on the physical quantities in the neighbourhood of the bouncing phase. Additionally, the Fig.~\ref{fig:p_rho_omega} [Bottom panel] shows the evolution of the EoS parameter $\omega=p/\rho$. It is observed that the Universe remains in the phantom regime ($\omega<-1$) around the bounce epoch. As cosmic time goes on, the EoS parameter crosses the $\Lambda$CDM boundary ($\omega=-1$) and evolves into the quintessence region ($-1<\omega<-1/3$), showing a change from a period of rapid acceleration to one of comparatively milder accelerated expansion.

As discussed in the previous section, the intended bouncing model is intrinsically non-symmetric with regard to cosmic time. This asymmetry is clearly reflected in the evolution of the EoS parameter. Around the bounce, the EoS exhibits phantom-like behaviour, while at later times it approaches the $\Lambda$CDM limit and eventually remains very close to $\omega=-1$. In contrast, during the pre-bounce phase, the EoS parameter evolves towards comparatively larger values, leading to a distinctly asymmetric evolution before and after the bounce. This asymmetric behaviour demonstrates that the background dynamics of the model differ significantly on either side of the bounce and are sensitive to the underlying non-metricity contributions.

\section{Energy Condition}\label{sec.5}

To determine the stability of the bouncing solution, it is essential to check the energy condition. In the literature, it is noted that the NEC is violated, and violation of the NEC implies violation of the pointwise energy conditions. We can express the general form of energy conditions in the context of $f(Q,T)=\alpha Q^m+\beta T$ \cite{AGRAWAL2021100863} gravity as,
\begin{table}[ht]
\centering
\caption{Energy conditions for a perfect fluid.}
\label{tab:EC}
\renewcommand{\arraystretch}{1.4}
\begin{tabular}{|c|c|c|}
\hline
\textbf{Condition} &
\textbf{Tensor Form} &
\textbf{Perfect Fluid Form} \\ \hline

NEC &
$T_{\mu\nu}k^\mu k^\nu\ge0$ &
$\rho+p\ge0$
\\ \hline

WEC &
$T_{\mu\nu}v^\mu v^\nu\ge0$ &
$\rho\ge0,\ \rho+p\ge0$
\\ \hline

SEC &
$\left(T_{\mu\nu}-\dfrac12 Tg_{\mu\nu}\right)v^\mu v^\nu\ge0$ &
$\rho+3p\ge0,\ \rho+p\ge0$
\\ \hline

DEC &
$T_{\mu\nu}v^\mu v^\nu\ge0$ &
$\rho\ge0,\ \rho\pm p\ge0$
\\ \hline

\end{tabular}
\end{table}

\begin{subequations}
\begin{eqnarray}
\rho + p &=& \frac{\Xi}{4\pi}\left(\frac{1}{1+\kappa}\right), \label{eq.13a}\\
\rho+3p&=&\frac{1}{16 \pi(1+2\kappa)(1+\kappa)}\left[-2(1-2m)(1+\kappa)\alpha Q^m+2\Xi (6+10\kappa)\right], \label{eq.13b}\\
\rho-p&=&\frac{1}{8\pi(1+2\kappa)}\left[(1-2m)\alpha Q^m-2\Xi\right].
\end{eqnarray}
\end{subequations}

\begin{figure}[H]
    \centering
    \includegraphics[width=0.45\textwidth]{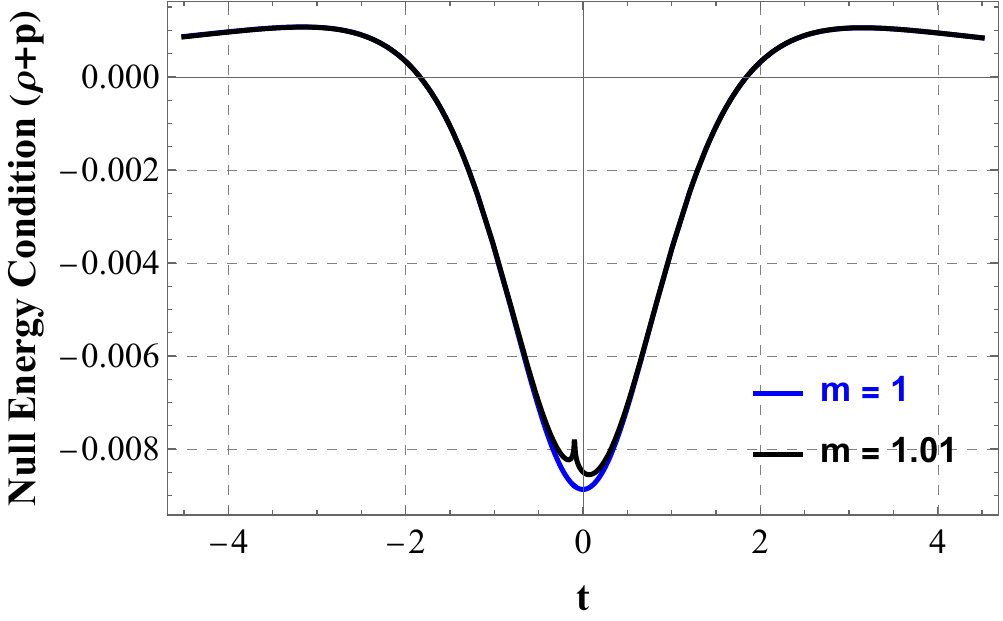}
    \hfill
    \includegraphics[width=0.45\textwidth]{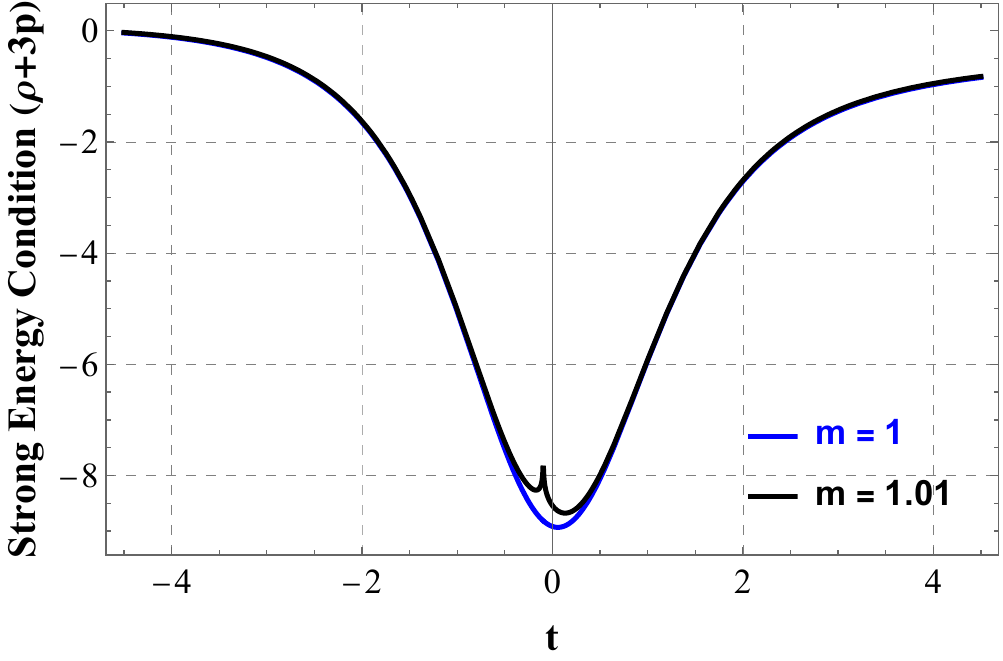}

    \includegraphics[width=0.45\textwidth]{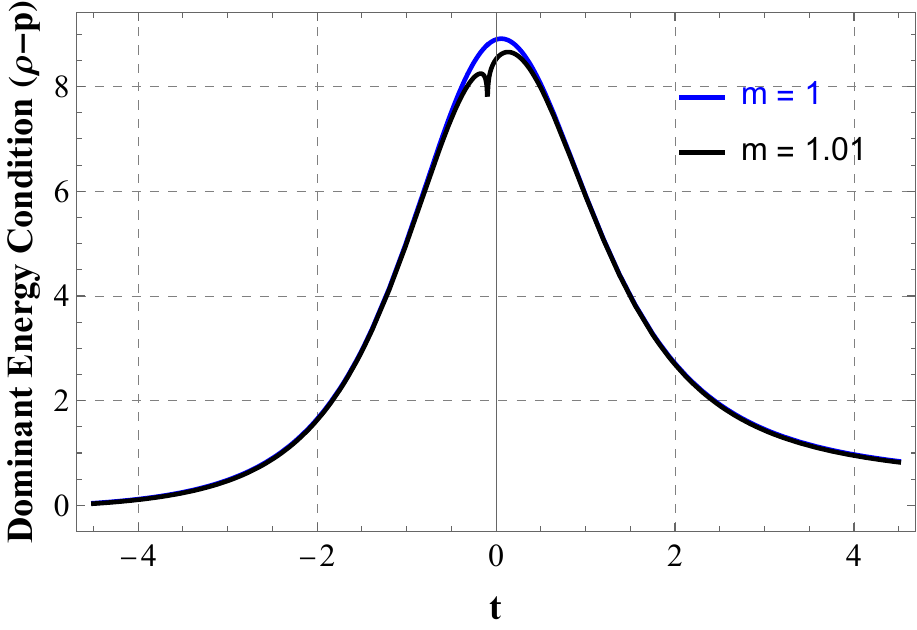}
    \caption{The evolution of the NEC [left] , SEC [right] and  DEC [bottom] are illustrated. The blue curve corresponds to the linear model with $m=1$, while the black curve represents the nonlinear case with $m=1.01$. The remaining model parameters are fixed as $\kappa=-0.49975$, $\alpha=-0.5$, $a_0=0.3$, $\gamma=\frac{4}{3}$, $n=0.185$, and $t_s=30$.}
    \label{fig:nec_sec}
\end{figure}

The NEC, represented by $ \rho+p $ (Left Panel Fig.~\ref{fig:nec_sec}), remains approximately zero throughout the considered range of $t$, indicating a marginal satisfaction of the NEC, but near the bounce epoch, we observed that it is violated. The SEC, represented by $\rho+3p$ (Right Panel Fig.~\ref{fig:nec_sec}), remains negative for all values of $t$, reaching its minimum near the bounce epoch. This persistent violation of the SEC suggests the presence of accelerated expansion or repulsive gravitational effects within the model. In contrast, the DEC, represented by $\rho-p$ (Bottom Fig.~\ref{fig:nec_sec}), remains positive over the entire domain, attaining a maximum away from the centre and exhibiting a shallow cusp near the bounce epoch. The positivity of the DEC indicates that the energy density dominates over the pressure, ensuring physically reasonable energy propagation. Overall, the model satisfies the dominant energy condition, marginally satisfies the null energy condition, and violates the strong energy condition, a behaviour commonly associated with cosmological models exhibiting accelerated expansion.

\section{Conclusion}\label{sec.6}

Considering the functional form $f(Q,T)=\alpha Q^{m}+\beta T$ in this work,where $\alpha$, $\beta$, and $m$ are free model parameters, we have investigated a non-singular bouncing cosmological solution in the framework of the $f(Q,T)$ gravity.
By adopting a suitable hybrid scale factor, we reconstructed a cosmological model that naturally unifies the early-time non-singular bounce with the late-time accelerated expansion of the Universe. The cosmological evolution of the model has been examined through the behaviour of the scale factor, Hubble parameter, and effective EoS parameter. The analysis shows that the Universe undergoes a smooth bounce at approximately $t\simeq -0.09$, where the Hubble parameter changes its sign from negative to positive, confirming the transition from a contracting to an expanding phase without encountering the initial singularity. Furthermore, the effective EoS parameter evolves consistently toward the present epoch and attains the value $\omega_{\rm eff}\simeq -1$ at $t=13.8~\mathrm{Gyr}$, in agreement with the observed dark energy-dominated Universe. We further derived the modified Friedmann equations and obtained analytical expressions for the pressure, energy density, and EoS parameter within the $f(Q,T)$ framework. The adopted hybrid scale factor combines the desirable features of the matter bounce scenario, which removes the initial singularity, with an exponential expansion responsible for the present cosmic acceleration. The resulting dynamical quantities exhibit characteristic bump/ditch behaviour around the bounce, reflecting the influence of the non-metricity correction governed by the parameter $m$. The analysis demonstrates that suitable choices of the model parameters yield a physically viable cosmic evolution while preserving the desired bouncing behaviour.

The energy conditions were also investigated to examine the physical consistency of the model. We found that the NEC is violated only in a small neighbourhood of the bounce, as required for a successful non-singular bounce, whereas the SEC and DEC exhibit behaviour consistent with the subsequent cosmological evolution. These results indicate that the bounce is realised through the modified geometric sector of the theory rather than by introducing exotic matter fields. Overall, the proposed $f(Q,T)=\alpha Q^{m}+\beta T$ model provides a self-consistent cosmological framework capable of describing the complete evolutionary history of the Universe. It successfully resolves the initial Big Bang singularity through a non-singular bounce while naturally evolving toward the observed late-time accelerated expansion. The compatibility of the model with both theoretical requirements and current cosmological observations suggests that $f(Q,T)$ gravity offers a promising alternative to the standard cosmological paradigm. Future investigations may focus on constraining the model parameters using observational datasets such as Type Ia supernovae, cosmic chronometers, BAO, and CMB measurements, as well as studying cosmological perturbations and structure formation to further assess the viability of the model.

\section*{Reference}
\bibliographystyle{unsrt}

\end{document}